\documentclass[12pt]{wlscirep}
\usepackage[utf8]{inputenc}
\usepackage[T1]{fontenc}
\usepackage{bm}
\usepackage{geometry}
\usepackage[english]{babel}
\usepackage{mathptmx}
\begin{document}
\title{\textbf{\Large{Growth morphology and symmetry selection of interfacial instabilities in anisotropic environments
}}}

\small{\author[1]{\fontfamily{ptm}\selectfont{Qing Zhang,}}
\author[2]{\fontfamily{ptm}\selectfont Amin Amooie,}
\author[2,3]{~\fontfamily{ptm}\selectfont Martin Z. Bazant,}
\author[1,*]{~\fontfamily{ptm}\selectfont Irmgard Bischofberger}}
\affil[1]{\fontfamily{ptm}\selectfont Department of Mechanical Engineering, Massachusetts Institute of Technology, Cambridge, MA 02139, USA}
\affil[2]{\fontfamily{ptm}\selectfont Department of Chemical Engineering, Massachusetts Institute of Technology, Cambridge, MA 02139, USA}
\affil[3]{\fontfamily{ptm}\selectfont Department of Mathematics, Massachusetts Institute of Technology, Cambridge, MA 02139, USA}
\affil[*]{\fontfamily{ptm}\selectfont e-mail: irmgard@mit.edu}

\begin{abstract}
\fontfamily{ptm}\selectfont {The displacement of a fluid by another less viscous one in a quasi-two dimensional geometry typically leads to complex fingering patterns. In an isotropic system, dense-branching growth arises, which is characterized by repeated tip-splitting of evolving fingers. When anisotropy is present in the interfacial dynamics, the growth morphology changes to dendritic growth characterized by regular structures. We introduce anisotropy by engraving a six-fold symmetric lattice of channels on a {H}ele-{S}haw cell. We show that the morphology transition in miscible fluids depends not only on the previously reported degree of anisotropy set by the lattice topography, but also on the viscosity ratio between the two fluids,~$\eta_{\mathrm{in}}$/$\eta_{\mathrm{out}}$. Remarkably, $\eta_{\mathrm{in}}$/$\eta_{\mathrm{out}}$ and the degree of anisotropy also govern the global features of the dendritic patterns, inducing a systematic change from six-fold towards twelve-fold symmetric dendrites. Varying either control parameter provides a new method to tune the symmetry of complex patterns, which may also have relevance for analogous phenomena of gradient-driven interfacial dynamics, such as directional solidification or electrodeposition.}
\end{abstract}



\flushbottom
\maketitle

\thispagestyle{empty}

\textbf{\large{Pattern growth}} is ubiquitous in nature and leads to the formation of complex structures~\cite{Gallaire2017,Cheng2008, zenit2019some}. Many interfacial patterns can be grouped into two ‘essential shapes’ or morphologies: isotropic dense-branching growth and anisotropic dendritic growth. Dense-branching growth arises from repeated tip-splitting of the structures and leads to a ramified pattern with many branches~\cite{Ben-Jacob1986,Paterson1981}, controlled by the gradient-driven transport of mass, heat or charge to the interface. In contrast, anisotropic dendritic growth is characterized by protrusions that are stable towards tip-splitting and leads to more regular patterns with global symmetries~\cite{Langer1980, Langer1989,Couder1990,Ben-Jacob1983,Ben-Jacob1990,Yasuda2019}. Here, we show that dendritic patterns -- with tunable symmetry -- can arise when the growth occurs in anisotropic environments.

The phenomenon of viscous fingering has played an important role in elucidating the basic principles of these two types of growth~\cite{Orr1984,Homsy1987,Amar2005,Cinar2009}, as well as methods to control the resulting patterns~\cite{Al-Housseiny2012,Al2013,Mirzadeh2017,Gao2019,videbaek2019diffusion,suo2020fingering,parsa2020origin,rosti2020breakdown,morrow2019numerical,lu2020computational,bischofberger2016fluid,setu2013viscous}. Viscous fingers result from the Saffman-Taylor instability, when one fluid is displaced by another less viscous one in the quasi-two dimensional geometry of a {H}ele-{S}haw cell~\cite{Saffman1958,HELE-SHAW1898}. It has been shown that dendritic growth requires anisotropy in the interfacial dynamics~\cite{Langer1989,Ben-Jacob1985,Ben-Jacob1988,Horvth1987}. In its absence, dense branching is instead the generic mode of growth~\cite{Homsy1987}. Anisotropy fixes the tip of an advancing interface into a stable parabolic shape that prevents it from splitting~\cite{Langer1989,Rabaud1988,Almgren1993,Ignes-Mullol1996} and introduces global symmetries along preferred growth directions, which are also seen in discrete models of diffusion-limited aggregation on crystal lattices~\cite{meakin1986universality,kertesz1986diffusion,stepanov2001laplacian}. Experimentally, anisotropy can be introduced either externally in the growth environment or internally in one of the fluids. External anisotropy can be imposed by engraving ordered channels on one of the plates of a {H}ele-{S}haw cell, by using channels confined with elastic membranes or by placing air bubbles at the tips of growing fingers~\cite{Ben-Jacob1985,Rabaud1988,Couder1986,Zocchi1987,Ducloue2017,McCue2018,Juel2018}. Internal anisotropy can be induced by replacing one of the fluids with a liquid crystal in the nematic phase~\cite{Buka1986}.

Previous studies have considered a particular limit of the viscous-fingering instability; the limit where the viscosity ratio between the less-viscous inner fluid and the more-viscous outer fluid, $\eta_{\mathrm{in}}$/$\eta_{\mathrm{out}}$, is very low, which is typically the case when air or water displace a viscous liquid. The patterns are then characterized by one single growing length scale, the finger length. Under these conditions, experiments using a {H}ele-{S}haw cell with engraved ordered channels have identified the degree of anisotropy, defined as the ratio between the channel height $h$ and the plate spacing $b$, $h$/$b$, as a control parameter for the morphology transition from dense-branching to dendritic growth~\cite{Ben-Jacob1985};  dendritic structures form beyond a critical value of $h$/$b$. When the two fluids are miscible, the degree of anisotropy is the only control parameter for the morphology transition. In the case of two immiscible fluids, the capillary number sets the critical $h$/$b$ for the transition~\cite{Ben-Jacob1988,Decker1999}. For miscible fluids and for immiscible fluids at high capillary number, the dendrites directly reflect the underlying symmetry of the lattice; four-fold symmetric dendrites grow in a four-fold symmetric lattice, six-fold symmetric dendrites grow in a six-fold symmetric lattice~\cite{Ben-Jacob1990}. Dendrites grow in the direction of the channels, which are the regions of largest effective plate spacing within which the flow velocity is highest~\cite{Ben-Jacob1985,Ben-Jacob1988}. 

We here reveal how a previously unexplored control parameter, the ratio of the viscosities of the inner and the outer fluid, $\eta_{\mathrm{in}}$/$\eta_{\mathrm{out}}$, modifies both the morphology transition and, remarkably, the symmetry of the dendritic structures in miscible fluids in anisotropic environments. Recent studies in isotropic environments have identified the viscosity ratio as an important control parameter that governs not only the onset of the instability~\cite{rana2017interaction,rana2019influence,sharma2020control, Bischofberger2014, videbaek2020delayed}, but also the global features of the patterns introducing a second length scale, the radius of a central region of complete outer-fluid displacement that grows concomitantly with the fingers~\cite{Bischofberger2014,Bischofberger2015,Jackson2015,Anjos2017}. This central region becomes increasingly larger, and therefore the relative length of the fingers increasingly smaller, as the viscosity ratio between the two fluids increases. Here we show that a morphology transition from dense-branching to dendritic growth can occur over a large range of viscosity ratios. We engrave channels creating a six-fold symmetric lattice on one of the plates and show that the critical degree of anisotropy, $h$/$b$, required for the transition to dendritic growth depends on the viscosity ratio between the two liquids. Remarkably, the dendrites can adopt a rich variety of emergent structures: they exhibit six-fold symmetric growth far from the morphology boundary and systematically transition towards twelve-fold symmetric structures as the boundary is approached. Our study reveals novel ways to tune both the morphology transition and the symmetry of dendritic patterns by either controlling the viscosity ratio between the two fluids or the geometric features of the growth environment.

\section*{\label{sec:level1}Methods}
Our experiments are performed in a radial {H}ele-{S}haw cell consisting of two 19~mm thick circular glass plates of diameter 280~mm. Six-fold symmetric lattices of diameter 145~mm are engraved on acrylic plates with a laser cutter (Universal Laser Systems) and placed on the bottom glass plate of the {H}ele-{S}haw cell. The width of the lattice channels $w$ and the distance between the edges of two channels $d$ are fixed to $w$~=~800~$\mu$m and $d$~=~850~$\mu$m (Fig.~1a). Four channel depths $h$ of 10~$\mu$m, 28~$\mu$m, 50~$\mu$m, and 250~$\mu$m are used. The plate spacing between the engraved acrylic plate and the top glass plate, $b$, is maintained by six spacers around the perimeter and varies from 125~$\mu$m to 1350~$\mu$m. The ratio between the height of the channel and the plate spacing, $h$/$b$, defines the degree of anisotropy. 

The miscible fluids used in our study are glycerol~(PTI Process Chemicals) and water~(VWR). We tune the viscosity of the inner fluid by mixing glycerol and water in different proportions and we use pure glycerol as the outer fluid. Details on the composition of the water-glycerol mixtures and their viscosities are reported in Table~S1 of the ESI. The fluids are injected through a 2~mm diameter hole in the center of one of the plates at a precise volumetric flow rate set by a syringe pump (Harvard PHD 2000). We use flow rates of 1~ml/min and 10~ml/min, which allows us to probe an order of magnitude difference in flow rate while staying in the high P\'eclet number regime~(Pe~=~$Ub/D_{12}$, where $U$ is the fingertip velocity and $D_{12}$ is the inter-diffusion coefficient~\cite{videbaek2019diffusion}), here ranging between 2100 -- 45240. Within this regime, the inter-diffusion of the fluids is negligible so that the fluids remain separated by a well-defined interface~\cite{videbaek2019diffusion}. The patterns are recorded with either a Point Grey camera (Grasshopper 3 GS3-U3-91S6M) at frame rates up to 9~fps or a LUMIX GH5 camera at frame rates up to 60~fps.

\begin{figure*}[t]
\centering
\includegraphics[width=1\linewidth]{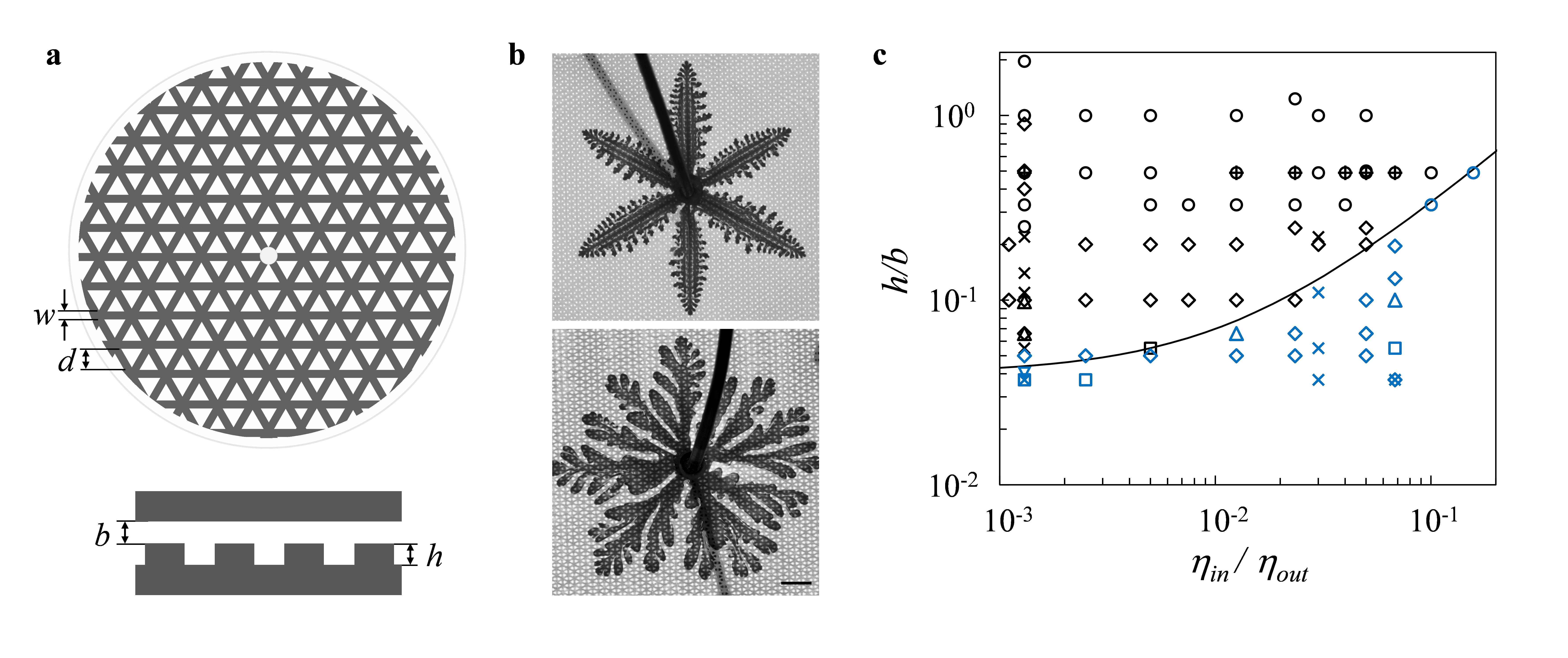}
\caption{(a) Schematic of the modified {H}ele-{S}haw cell. Top image: top view of the bottom plate of the {H}ele-{S}haw cell with an engraved six-fold symmetric lattice, with width of the lattice channels $w$ and distance between the edges of two channels $d$. Bottom image: side view of the modified {H}ele-{S}haw cell, denoting the plate spacing $b$ and the channel height $h$. (b) Examples of dendritic growth (top, for $h$/$b$~=~0.5, $h$~=~50~${\mu}$m, $b$~=~100~${\mu}$m) and dense-branching growth (bottom, for $h$/$b$~=~0.04, $h$~=~10~${\mu}$m, $b$~=~254~${\mu}$m) at low viscosity ratio $\eta_{\mathrm{in}}$/$\eta_{\mathrm{out}}$~=~0.0013. The scale bar is 1~cm. 
(c) Morphology diagram controlled by the viscosity ratio $\eta_{\mathrm{in}}$/$\eta_{\mathrm{out}}$ and the degree of anisotropy $h$/$b$. Blue symbols denote dense-branching growth, black symbols denote dendritic growth. Experiments are performed with engraved plates with different channel heights~$h$ and plate spacings~$b$ and at different volumetric flow rates~$q$. ({{${\triangledown}$}}) $h$~=~10~${\mu}$m, $q$~=~1ml/min; ($\pmb{\times}$) $h$~=~28~${\mu}$m, $q$~=~1ml/min; ({\scriptsize{$\pmb{\square}$}})~$h$~=~28~${\mu}$m, $q$~=~10~ml/min; ({\Large${\diamond}$}) $h$~=~50 ${\mu}$m, $q$~=~1 ml/min; ({\scriptsize{$\pmb{\triangle}$}}) $h$~=~50 ${\mu}$m, $q$~=~10~ml/min; ({\Large{${\circ}$}}) $h$~=~250 ${\mu}$m, $q$~=~1ml/min; (+) $h$~=~250~${\mu}$m, $q$~=~10 ml/min. The value of $b$ for each experiment is listed in Table S2 of the ESI. The solid line denotes a fit to ($h$/$b$~$-$~($h$/$b$)$^{\ast}$)~/~($\eta_{\mathrm{in}}$/$\eta_{\mathrm{out}}$)~=~$A$  ($A$~=~3 and ($h$/$b$)$^{\ast}$~=~0.04 are best-fit parameters).}
\label{fig:1}
\end{figure*}

We complement the experiments with two-dimensional (2D) high-resolution numerical simulations using the finite element software COMSOL Multiphysics (v5.4), which allows us to access the pressure distribution in the fluids. Our model replicates the geometry of the {H}ele-{S}haw cell in terms of the cell diameter and the six-fold symmetric lattice dimensions. The lowest viscosity ratio we can access in our simulations (${\eta_ {\mathrm{in}}}/{\eta _{\mathrm{out}}}~=~0.006$) is slightly higher than that probed in experiments (${\eta_ {\mathrm{in}}}/{\eta _{\mathrm{out}}}~=~0.0011$), as for very low ${\eta_ {\mathrm{in}}}/{\eta _{\mathrm{out}}}$ the fingertip velocity becomes too fast compared to the mean flow velocity, which results in numerically unstable solutions. This is a known issue for a number of numerical approaches~\cite{jha2011quantifying}, but one that could be overcome by designing numerical schemes suited for low viscosity ratios~\cite{islam2005fully}. It, however, does not prevent us from accessing the full range of patterns observed in the experiments.

We employ the finite element method to solve the partial differential equations. We couple the convection-diffusion mass-transport equation from the \emph{Transport of Diluted Species} Module with the continuity equation for the single-phase, incompressible flow velocity from the \emph{Darcy’s Law} Module. The governing equations are:
\begin{eqnarray}
\frac{{\partial c}}{{\partial t}} + \nabla  \cdot \left( {- D\nabla c + c{\mathbf{u}}} \right) = 0
\end{eqnarray}
\begin{eqnarray}
{\mathbf{u}} =  - \frac{k}{\eta}\nabla p
\end{eqnarray}
\begin{eqnarray}
\nabla \cdot {\mathbf{u}} = 0
\end{eqnarray}
where $c$ is the concentration of the inner fluid and $D$ the molecular diffusion coefficient. The latter is chosen as $D~=~{10^{-14}}$~m$^{2}$/s given the high P{\'e}clet numbers of the experiments. We note that the pattern morphology remains independent of $D$ for $D~<~10^{-8}$~m$^{2}$/s, confirming that our simulations are in the high P{\'e}clet number regime (see ESI for further details). ${\nabla}$ is the in-plane gradient operator, ${\mathbf{u}}$ is the Darcy velocity set by the pressure gradient ${\nabla p}$, and $k$ and ${\eta}$ are the permeability and viscosity of the fluids, respectively. 

The flow in a {H}ele-{S}haw cell can be approximated as quasi-2D as the plate spacing, $b$, is much smaller than the radial dimension. The gap-averaged velocity of the fluids is then
${\mathbf{u}}~=~-\frac{b_{i}^2}{12\eta}\nabla p$ with ${k~=~b_{i}}^2/12$, where $b_{i}$ describes the gap thickness at any point of the textured surface. The spatial variability in the plate spacing is incorporated in the numerical model by defining a binary spatial distribution of permeability~[L${^{2}}$], consisting of a permeability value for the obstacles (assigned to the triangles forming the lattice cells), denoted as $k_{1}$, and a permeability value for the channels (assigned to the background domain), denoted as $k_{2}$. The ratio between the two permeabilities, $k_{2}$/$k_{1}$, is~${\left(1+ h/b \right)^2}$. We use an exponential mixing rule for the mixture viscosity $\eta$ and $\eta~=~\eta _{\mathrm{out}}{e^{ - M c}}$, where $\eta _{\mathrm{out}}$ is the viscosity of the outer fluid and ${\eta_ {\mathrm{in}}}/{\eta _{\mathrm{out}}}~=~{e^{ - M}}$. For miscible fluids, both the pressure and the normal velocity are continuous at the interface. We define a small circular inlet region around the cell center which provides a smooth boundary, to avoid a point-source injection that could lead to a singularity in the domain. A normal inflow velocity for flow and a Dirichlet boundary condition ($c$~=~1) for transport are applied at the perimeter of the circular inlet region, and atmospheric pressure (open-flow) condition for flow and an outflow condition (${\mathbf{n}}\cdot D\nabla c~=~0$) for transport are imposed on the outer cell boundary. The initial conditions in the entire domain are $c$~=~0 and $p$~=~0. The absolute values of the injection velocity and the permeability within the computational domain differ from those in experiments. This does not affect the resulting patterns, as only the ratio of the permeabilities ($\sim h/b$) governs the pattern morphology. 

We solve for pressure and concentration fields in a fully coupled approach using the Parallel Direct Sparse Solver Interface (PARDISO)  and Newton's method with dynamic damping for highly nonlinear systems. The implicit Generalized-$\alpha$ Method is used for the time stepping scheme \cite{Chung1993,jansen2000generalized}. We use the default discretization settings that govern the order of discretization in the shape functions for the dependent variables of each module: first-order discretization for the convection-diffusion equation and second-order discretization for Darcy's law, as these settings work efficiently and robustly. The optimal mesh resolution is found with these discretization orders. The annular mesh area used is 0.00606 m$^{2}$ discretized by 222,162 triangular elements. We have confirmed the numerical validity and convergence of our simulations (see ESI). The discretization by a triangular mesh provides a source of perturbation sufficient for the instability to occur; the apparent slight asymmetry of the computed patterns is mesh driven due to the spatial non-uniformity of the perturbation and the triangularization of the domain.  

\section*{\label{sec:level2}Results}
\subsection*{Morphology transitions of miscible viscous fingering in an anisotropic {H}ele-{S}haw cell}
We investigate the growth of patterns in anisotropic environments by engraving channels creating a six-fold symmetric lattice on one of the {H}ele-{S}haw plates, as shown in Fig.~1a. We use pairs of miscible fluids with different ratios of viscosities between the less-viscous inner fluid and the more-viscous outer fluid, $\eta_{\mathrm{in}}$/$\eta_{\mathrm{out}}$. The use of miscible fluids allows us to investigate the role of viscosity ratio without concurrently varying the capillary number.

In agreement with previous studies at very low viscosity ratios and high capillary numbers, we find that the morphology transition from dense-branching growth to dendritic growth occurs above a value of $h$/$b$ ${\approx 0.05}$~\cite{Ben-Jacob1985, Decker1999} for our lowest $\eta_{\mathrm{in}}$/$\eta_{\mathrm{out}}$. Below this value, fingers grow by repeated tip-splitting which results in dense-branching growth, above this value, the fingertip is stabilized which results in dendritic growth, as shown in Fig.~1b. Remarkably though, this critical  $h$/$b$ depends strongly on the viscosity ratio: as $\eta_{\mathrm{in}}$/$\eta_{\mathrm{out}}$ increases, a larger $h$/$b$ is needed to transition from dense-branching growth to dendritic growth, as shown in Fig.~1c. 
We find that neither the absolute values of the channel height $h$ and the plate spacing $b$, nor the volumetric flow rate are control parameters for the morphology transition, as shown by the different symbols in Fig.~1c, which denote experiments performed with plates of various channel heights $h$ ranging from 10~$\mu$m to 250~$\mu$m, various plate spacings $b$ ranging from 125~$\mu$m to 1350~$\mu$m, and at two volumetric flow rates of 1~ml/min and 10~ml/min. For a given viscosity ratio, any combination of $h$ and $b$ yielding a certain value of $h$/$b$ leads to the same growth morphology. 

\begin{figure}[t]
\centering
\includegraphics[width=0.65\linewidth]{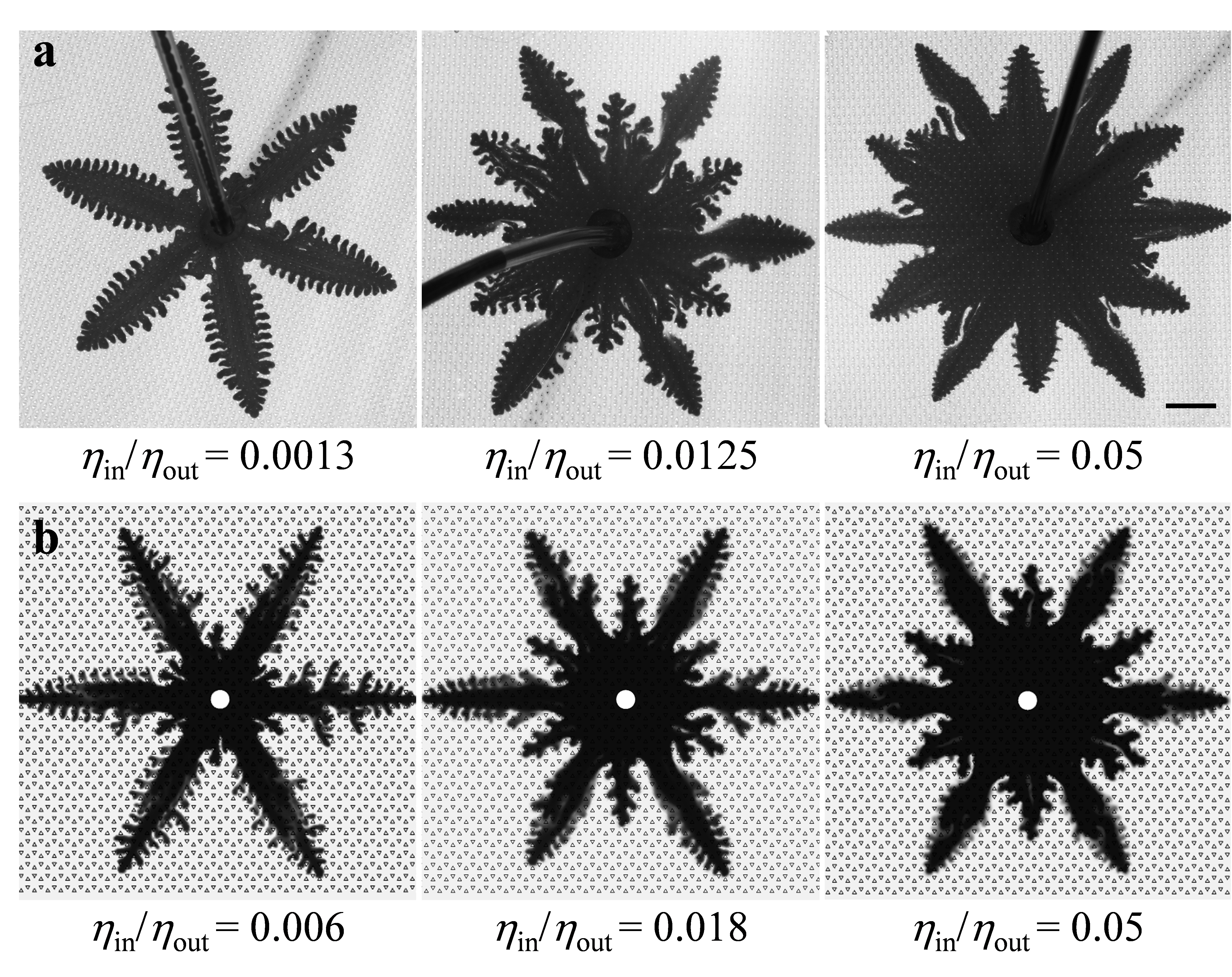}
\caption{{Systematic change from six- towards twelve-fold symmetric dendrites.} (a)  Dendritic patterns  for different viscosity ratios obtained at $h$/$b$~=~0.49. As $\eta_{\mathrm{in}}$/$\eta_{\mathrm{out}}$ increases, the additional generation of sub dendrites grows progressively larger. The scale bar is 1 cm. (b) Snapshots of the simulations at $h$/$b$~=~0.49.}
\label{fig:2}
\end{figure}

\subsection*{Dendritic growth adopts different symmetries}
The viscosity ratio $\eta_{\mathrm{in}}$/$\eta_{\mathrm{out}}$ and the degree of anisotropy $h$/$b$ not only determine the morphology boundary, but have a more dramatic effect on the pattern growth in the dendritic regime. For a fixed $h$/$b$, an increase in the viscosity ratio $\eta_{\mathrm{in}}$/$\eta_{\mathrm{out}}$ leads to a systematic change in the pattern symmetry. Remarkably, the imposed six-fold symmetry of the engraved plate leads to six-fold symmetric growth only at the lowest viscosity ratio. At higher viscosity ratios, the pattern instead transitions towards a twelve-fold symmetry; in addition to the six main dendrites evolving along the straight channels, additional six sub dendrites emerge at a 30$^{\circ}$ angle to the preferred growth direction, as shown in Fig.~2a. The length of the sub dendrites becomes larger with increasing viscosity ratio and eventually comparable to that of the main dendrites. A similar trend is recovered in the simulations, as shown in Fig.~2b. 
A transition from six- towards twelve-fold symmetry also occurs for a fixed $\eta_{\mathrm{in}}$/$\eta_{\mathrm{out}}$ with decreasing $h$/$b$ (see Fig.~S3 of the ESI). Previous studies in the limit of low viscosity ratios have seen hints towards the onset of these additional sub dendrites \cite{Chen1987,Banpurkar1999,Banpurkar2000}. Here we show their systematic growth and that they can become comparable in size to the main dendrites within a certain range of viscosity ratio and $h$/$b$. 

\begin{figure*}[h]
\centering
\includegraphics[width=1.\linewidth]{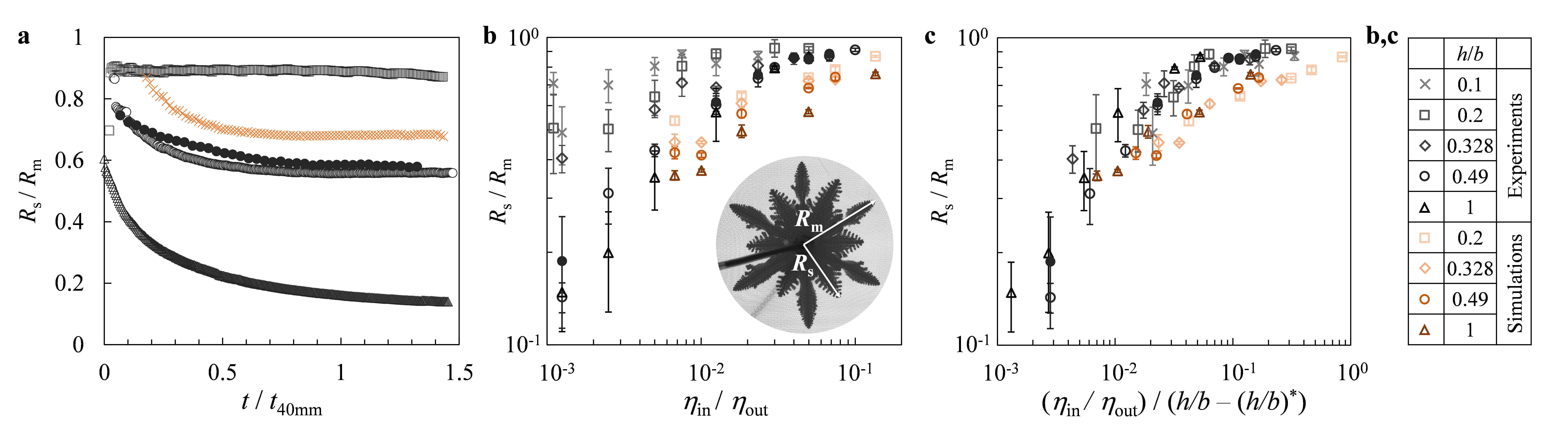}
\caption{(a) Temporal evolution of $R_{\mathrm{s}}$/$R_{\mathrm{m}}$ for $\eta_{\mathrm{in}}$/$\eta_{\mathrm{out}}$~=~0.0013, $h$/$b$~=~0.49 and $q$~=~1~ml/min~({\scriptsize{$\pmb{\triangle}$}}), $\eta_{\mathrm{in}}$/$\eta_{\mathrm{out}}$~=~0.0125, $h$/$b$~=~0.49 and $q$~=~1~ml/min~({\Large{${\circ}$}}), $\eta_{\mathrm{in}}$/$\eta_{\mathrm{out}}$~=~0.0125, $h$/$b$~=~0.49 and $q$~=~10 ml/min~({\Large${\bullet}$}), $\eta_{\mathrm{in}}$/$\eta_{\mathrm{out}}$~=~0.05, $h$/$b$~=~0.49 and $q$~=~1 ml/min~({\scriptsize{$\pmb{\square}$}}) in experiments, and for $\eta_{\mathrm{in}}$/$\eta_{\mathrm{out}}$~=~0.05, $h$/$b$~=~0.49~(${\textcolor{orange}{\times}}$) in simulations. $t_{\mathrm{40mm}}$ is the time when $R_{\mathrm{m}}~=~$~40~mm. (b) $R_{\mathrm{s}}$/$R_{\mathrm{m}}$ versus $\eta_{\mathrm{in}}$/$\eta_{\mathrm{out}}$ for different $h$/$b$ and $q$. $R_{\mathrm{s}}$/$R_{\mathrm{m}}$ is measured when $R_{\mathrm{m}}$~=~40~mm. The symbols are defined in the table. Open symbols denote $q$~=~1~ml/min, closed symbols denote $q$~=~10~ml/min. (c) Scaled master curve of $R_{\mathrm{s}}$/$R_{\mathrm{m}}$ versus ($\eta_{\mathrm{in}}$/$\eta_{\mathrm{out}}$)/($h$/$b-(h$/$b$)$^{\ast}$). The monotonic increase in $R_{\mathrm{s}}$/$R_{\mathrm{m}}$ denotes the change from six-fold towards twelve-fold symmetric dendritic patterns. The symbols are the same as in (b).}
\label{fig:3}
\end{figure*}

To quantify the change from six- towards twelve-fold symmetry, we define the length of the main dendrites, $R_{\mathrm{m}}$, corresponding to the structures growing in the direction of the six straight channels, and the length of the sub dendrites, $R_{\mathrm{s}}$, corresponding to the structures growing at an angle of 30$^{\circ}$ with respect to the six straight channels, as shown in the inset of Fig.~3b. The ratio $R_{\mathrm{s}}$/$R_{\mathrm{m}}$ exhibits a transient regime at early times and then remains almost constant in time for fully developed patterns. Moreover, $R_{\mathrm{s}}$/$R_{\mathrm{m}}$ is independent of the interfacial velocity for the range of flow rates investigated, as shown in Fig.~3a, where we normalize the time by $t_{\mathrm{40mm}}$ denoting the time when $R_{\mathrm{m}}~=~40$~mm. To compare the patterns formed at different viscosity ratios,
we measure $R_{\mathrm{s}}$/$R_{\mathrm{m}}$ when $R_{\mathrm{m}}~=~40$~mm, which is well within the fully developed regime. For a fixed $h$/$b$, the ratio $R_{\mathrm{s}}$/$R_{\mathrm{m}}$ monotonically increases with viscosity ratio. In addition, a decrease in $h$/$b$ leads to an increase in $R_{\mathrm{s}}$/$R_{\mathrm{m}}$, as shown in Fig.~3b. We can rescale all data by normalizing the viscosity ratio with ($h$/$b$~$-$~($h$/$b$)$^{\ast}$), as shown in Fig.~3c. The factor ($h$/$b$)$^{\ast}$ will become evident in the discussion of the morphology boundary. The numerical results are in good qualitative agreement with the experiments and exhibit the same scaling with $h$/$b$, but yield slightly lower values of $R_{\mathrm{s}}$/$R_{\mathrm{m}}$ compared to the experimental results. This is likely due to the 2D nature of the simulations (where we average the flow in the third dimension across the gap and assume a parabolic velocity profile in the gap direction~\cite{batchlor1967introduction, saffman1986viscous, gondret1997viscous}), which do not capture effects related to the partial displacement of the outer fluid or to the three-dimensional tongue-like structures that form between miscible fluids in a Hele-Shaw cell~\cite{lajeunesse19973d,lajeunesse1999miscible,Bischofberger2014}. Exploring further improvements to the model, e.g., solving Stokes flow in the full 3D domain, and a deeper investigation into quantitative comparisons with experiments are interesting topics for future work.

\section*{\label{sec:level5}Discussion}
The observation that both $\eta_{\mathrm{in}}$/$\eta_{\mathrm{out}}$ and $h$/$b$ allow one to systematically tune the symmetry of the patterns reveals a novel aspect of dendritic growth. Remarkably, the change in symmetry is also directly linked to the morphology transition to dense-branching growth: When $R_{\mathrm{s}}$/$R_{\mathrm{m}}$ reaches ${\sim0.85}$, corresponding to patterns with twelve dendrites of almost equal size, a further decrease in $h$/$b$ or a further increase in $\eta_{\mathrm{in}}$/$\eta_{\mathrm{out}}$ induces the transition to dense-branching growth. The morphology transition can therefore be described by the same functional form used to normalize the data in Fig.~3c; the morphology boundary denoted by a solid line in Fig.~1c corresponds to
\begin{equation}
\frac{h}{b}~=~A\frac{\eta_{in}}{\eta_{out}}+\left({\frac{h}{b}} \right)^{\ast} 
\label{eq:one}
\end{equation}
where $A$~=~3 and $(h/b)^{\ast}$~=~0.04 are best-fit parameters determined by logistic regression. $(h/b)^{\ast}$ denotes the critical $h$/$b$ for the morphology transition in the limit of low viscosity ratio.

\begin{figure*}[t]
\centering
\includegraphics[width=0.9\linewidth]{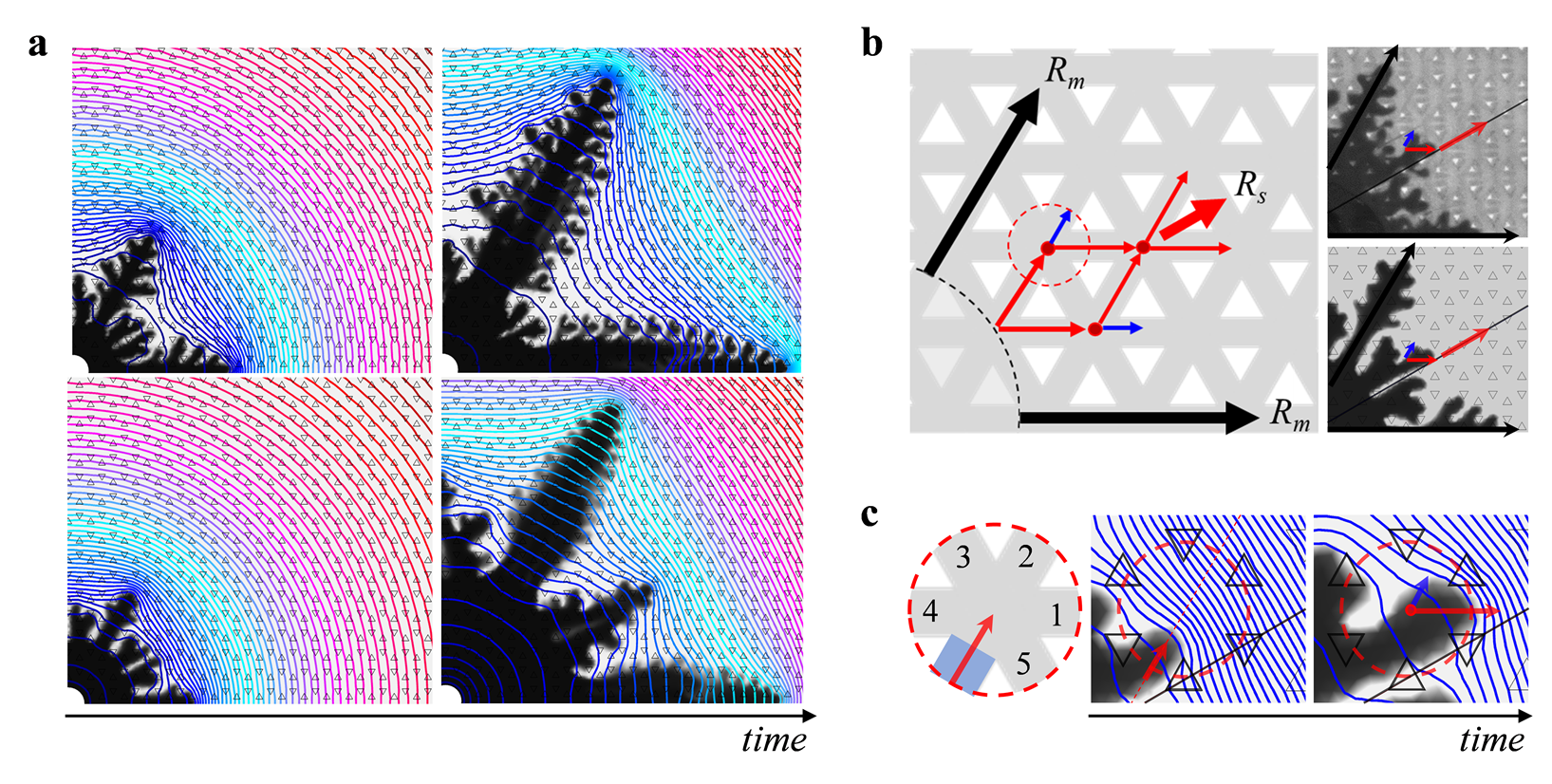}
\caption{{Formation and growth of sub dendrites.} (a) Pressure field for patterns with $\eta_{\mathrm{in}}$/$\eta_{\mathrm{out}}$~=~0.006 (top) and $\eta_{\mathrm{in}}$/$\eta_{\mathrm{out}}$~=~0.05 (bottom) at two different times. The lines indicate pressure contours. The pressure field ranges from atmospheric pressure (denoted by red contours) to a maximum pressure around the inlet (denoted by blue contours), which is different for each panel as it varies with time and viscosity ratio; the colors are guides to the eye. (b) Schematic representation of the path followed by the main dendrites $R_{\mathrm{m}}$ and the sub dendrites $R_{\mathrm{s}}$. At a lattice junction (indicated by the red dot in the dotted circle), the flow predominantly selects the direction along the red arrow, which leads to the growth of the sub dendrites along the 30$^{\circ}$ direction, as observed in both experiment (top image) and simulation (bottom image). (c) Zoomed schematics of the lattice junction. The combination of the global pressure distribution from the main dendrites and the local pressure distribution from the tip of the sub dendrites leads to flow into channel 1 along the direction of the red arrow.}
\label{fig:4}
\end{figure*}

Why do six-fold dendritic patterns only form far from the morphology boundary, and what leads to the growth of an additional generation of dendrites as we approach the boundary? The importance of the viscosity ratio and the degree of anisotropy for determining ${R_{\mathrm{s}}/R_{\mathrm{m}}}$ can be seen in a simplified analysis taking into account the effective permeability at different locations corresponding to the growth of sub dendrites or main dendrites, as detailed in the ESI. Note that the effective permeability in our system is isotropic and lacks a macroscopic preferred direction for single-phase flow. In general, the permeability tensor must be symmetric (by Onsager reciprocity for Stokes flow) and positive definite (by the Second Law of Thermodynamics) and thus represented by an orthogonal matrix~\cite{torquato2013}, so its eigenvectors, corresponding to the fastest and slowest directions, must be mutually perpendicular~\cite{bazant2008}. This orthogonality is incompatible with triangular symmetry, thus the permeability eigenvalues in our textured Hele-Shaw cell must be degenerate, implying isotropic single-phase flow. 

For two-phase flow, however, the gradient of viscosity at the interface between the two fluids can locally break the symmetry and induce an anisotropic effective permeability near the interface. Using concepts derived for the hydrodynamics of slippage on textured surfaces for two-phase flows over hydrophobic surfaces~\cite{Vinogradova1999, feuillebois2009}, we consider that the more-viscous outer fluid is partially trapped in the texture as the tip of the less-viscous fluid passes over the texture in the middle of the channel along the ``path of least resistance''. For small textures, the trapped fluid leads to a local effective slip length tensor~\cite{stroock2002,stone2004}, ${{\mathbf{b}}_{\mathrm{slip}}}$, which causes the effective permeability tensor to become anisotropic and orthogonal in the vicinity of the interface~\cite{bazant2008}, leading to the appearance of sub dendrites that impart this square symmetry to the pattern. In the limit of ``weak anisotropy'' in the slip tensor, $\mathrm{Tr}({{\mathbf{b}}_{\mathrm{slip}}})~<<~b$, as is the case for our experiments, we find that the interface velocities of the sub dendrites and the main dendrites, and therefore ${R_{\mathrm{s}}/R_{\mathrm{m}}}$, are indeed governed by $\eta_{\mathrm{in}}$/$\eta_{\mathrm{out}}$ and ${h/b}$. 

To get further insight into the growth of the dendrites, we consider their macroscopic path selection. The main dendrites $R_{\mathrm{m}}$ grow along the six straight channels. The sub dendrites $R_{\mathrm{s}}$ select a path at a 30$^{\circ}$ angle from these straight channels. At early stage, two fingers form between each pair of neighboring main dendrites on each side of the 30$^{\circ}$ direction, due to the anisotropy of the lattice. This is observed at any viscosity ratio, as shown in Fig.~4a. Whether these fingers will merge towards each other and grow into a sub dendrite or merge with the main dendrites resulting in a six-fold symmetric pattern depends on the pressure distribution imposed both globally by the main dendrites and locally at the tip of the sub dendrites. At low $\eta_{\mathrm{in}}$/$\eta_{\mathrm{out}}$ and high $h$/$b$, the rapid growth of the main dendrites sets up a large pressure gradient at their tip which in turn induces a small pressure gradient in the 30$^{\circ}$ direction, as shown in Fig.~4a, which prevents the sub dendrites from growing. With increasing $\eta_{\mathrm{in}}$/$\eta_{\mathrm{out}}$ and decreasing $h$/$b$, however, the sub dendrites themselves build locally a high pressure gradient at their tips which amplifies their growth. We provide further details on the growth of the sub dendrites in Figs. S5 and S6 of the ESI.

Once the sub dendrites have emerged, they continue to grow along the 30$^{\circ}$ direction following a zig-zag path, as illustrated in Fig.~4b. As the tip of the sub dendrite reaches a lattice junction, indicated by a red dot, the path towards the 30$^{\circ}$ direction (red arrow) is selected, rather than the straight path (blue arrow). 
This selection results from the pressure profile induced in the outer fluid by the main dendrites, which effectively shields the sub dendrites from growing towards the main dendrites and pushes them towards the 30$^{\circ}$ direction. Indeed, when the tip of a sub dendrite reaches the entrance of a lattice junction, as schematically shown in the zoomed-in region in Fig.~4c, it does not grow straight towards channel~2, but is deviated towards the 30$^{\circ}$ direction as a result of the global pressure distribution built up by the neighboring main dendrites. The local pressure distribution at the tip of the sub dendrite then induces a maximum pressure gradient towards channel~1, and most of the flow goes into channel~1. It is this combination of the global pressure distribution from the main dendrites and the local pressure distribution from the tip of the sub dendrites that leads to the rich pattern selection in dendritic growth.

These different paths selected by the main dendrites and sub dendrites also reveal the origin of the maximum value of $R_{\mathrm{s}}/R_{\mathrm{m}}~\approx~0.85$. It reflects the condition where the velocity of the main and sub dendrites becomes approximately equal. As the path selected by the sub dendrites deviates from the radial direction at each junction, the total path is 2/${\sqrt{3}}$ times longer than that of the main dendrites in the straight radial channels. The length of the main dendrite, $R_{\mathrm{m}}$, is therefore $(2/\sqrt {3}){R_{\mathrm{s}}}$, i.e., $R_{\mathrm{s}}/{\rm{ }}R_{\mathrm{m}} \sim \sqrt{3}/2{\rm{ }}~=~{\rm{ }}0.866$.
Interestingly, our experiments show that once this condition is reached, a further increase in $\eta_{\mathrm{in}}$/$\eta_{\mathrm{out}}$ or decrease in $h$/$b$ induces the morphology transition to dense-branching growth. This suggests that the morphology transition occurs when the difference between the pressure gradient in the straight channels and the 30$^{\circ}$ direction becomes negligible, and therefore the role of anisotropy becomes negligible, such that the parabolic tips can no longer be stabilized allowing for tip-splitting to occur. Reaching a full understanding of this morphology transition can be topic of further research.

\section*{\label{sec:level6}Conclusions}
Our results reveal a rich morphology of patterns created by pairs of miscible fluids in anisotropic systems. They demonstrate the important role of the viscosity ratio between the two fluids, which, together with the degree of anisotropy, governs both the morphology transition from dense-branching to dendritic growth and the selected symmetry of the dendrites. Upon approaching the morphology boundary, the dendritic patterns systematically transition from six-fold towards twelve-fold symmetry in the parameter regime where interfacial flow is governed by an effective slip tensor, whose orthogonality imparts square symmetry to the original pattern. 

This diversity of different dendritic patterns provides novel opportunities for tuning the growth of complex structures, not only in viscous fingering, but perhaps also in other cases of interfacial motion limited by gradient-driven transport processes, which lie in the same universality class~\cite{bazant2003dynamics}. In general, we expect that dendritic growth following the preferred directions of an anisotropic environment will tend to acquire orthogonal symmetry for ``weak anisotropy'', whenever transport near the interface is governed by a local effective conductance tensor, which must be orthogonal like the effective slip tensor in a weakly textured Hele-Shaw cell~\cite{bazant2008}. For example, in template-assisted directional solidification~\cite{de2000microdomain,liu2013templated}, a similar morphological transition may arise, controlled by the ratio of thermal diffusivities (analogous to the ratio of inverse viscosities here), whenever the pattern is controlled by the conduction of latent heat away from the interface in the liquid phase. Similarly, in template-assisted electrodeposition~\cite{bera2004synthesis,liu2008fabrication,han2014over,han2016dendrite}, it may be possible to tune the symmetry of dendritic patterns by varying the strength of diffusion anisotropy in the electrolyte domain.  Active control of anisotropic dendritic growth may also be achieved, for example, by applying electric fields to control viscous fingering~\cite{Mirzadeh2017,Gao2019} over patterned, charged surfaces~\cite{ajdari2001transverse} having anisotropic electro-osmotic slip tensors~\cite{bahga2010anisotropic}.

\section*{Acknowledgements}
We thank Thomas E. Videb\ae k, Pedro Saenz and Anne Juel for helpful discussions. Q.Z. and I.B. acknowledge support from the MIT Research Support Committee.


\section*{Conflicts of interest}
There are no conflicts to declare.

\bibliography{sample}

\end{document}